\documentclass[doublecol]{epl2}
\usepackage{graphicx}
\usepackage{amsmath}
\usepackage{bbm}

\def\eq#1{eq.~(\ref{#1})}
\def\fig#1{fig.~\ref{#1}}

\newcommand{\de}{\mbox{d}}

\newcommand{\mean}[1]{\langle #1 \rangle}

\AtBeginDocument{\bibliographystyle{eplbib}}

\begin{document}
\title{Mean encounter times for cell adhesion in hydrodynamic flow: analytical progress by dimensional reduction}
\author{C. B. Korn \and U. S. Schwarz}
\shorttitle{Cell adhesion in hydrodynamic flow}
\shortauthor{C.B. Korn \etal}
\institute{
  University of Heidelberg, Bioquant 0013, %
   Im Neuenheimer Feld 267, D-69120 Heidelberg, Germany
}

\abstract{For a cell moving in hydrodynamic flow above a wall,
  translational and rotational degrees of freedom are coupled by the
  Stokes equation. In addition, there is a close coupling of
  convection and diffusion due to the position-dependent mobility.
  These couplings render calculation of the mean encounter time
  between cell surface receptors and ligands on the substrate very
  difficult. Here we show for a two-dimensional model system how
  analytical progress can be achieved by treating motion in the
  vertical direction by an effective reaction term in the mean first
  passage time equation for the rotational degree of freedom.  The
  strength of this reaction term can either be estimated from
  equilibrium considerations or used as a fit parameter.  Our
  analytical results are confirmed by computer simulations and allow
  to assess the relative roles of convection and diffusion for
  different scaling regimes of interest.}

\maketitle

\section{Introduction}

Biological function is often based on the formation of a specific
binding complex between receptor and ligand \cite{alberts}. However,
in order for binding to occur, a physical transport process must exist
which brings the binding partners to sufficiently close proximity
\cite{berg:77}. In many cases of interest, this transport process is
rather complex. Usually it contains several coupled degrees of
freedom, like a cell surface receptor moving laterally on a membrane
which fluctuates in the vertical direction \cite{seifert:05c}. The
efficiency of biological transport processes often can be framed as
mean first passage time (MFPT) problems, for example for the gating of
ion channels \cite{haenggi:02} or the arrival of a virus at the
nucleus \cite{holcman:07}. Another example of a complex transport
process of large biological relevance is the receptor-mediated
adhesion of cells which are carried over a ligand-coated substrate by
hydrodynamic flow \cite{lawrence:91}. Here the mean encounter time
between receptors and ligands is a measure for the efficiency of cell
adhesion under the conditions of hydrodynamic flow \cite{korn:06}.
For this system, additional complications arise from the presence of
multiple length scales. For the micron-sized cell, the hydrodynamic
equations result in coupling of the translational and rotational
degrees of freedom.  Even for high shear rates, Brownian motion is
relevant because receptors and ligands are nanometer-sized objects,
thus even small movements for the cell result in a large effect on the
molecular level.  Here we show that despite the presence of these
complications, analytical progress can be achieved by dimensional
reduction of appropriate degrees of freedom. The reduced description
then contains effective parameters which have to be obtained from the
full model.

Experimentally the binding of cells to a substrate in hydrodynamic
flow is often studied in flow chambers because this setup allows for
controlled flow conditions \cite{alon:97,springer:01}. In vivo, this
situation is relevant for white blood cells, which travel the body
with the blood flow, but have to adhere at very specific locations,
e.g.\ close to sites of inflammation. Similar mechanisms are used by
cancer and stem cells.  Moreover, malaria-infected red blood cells
also undergo adhesion to the vessel walls under flow conditions. Apart
from using flow chambers, one can further reduce the experimental
system by employing biomimetic analogs of cells, that is receptor
bearing micro-beads \cite{pierres:98,hammer:00a}. The efficiency with
which cells or beads in flow can bind to a substrate depends crucially
on the spatial distribution of receptors and ligands\cite{yago:07}.
Previously, we proposed a model based on a Langevin equation for a
spherical particle in linear shear flow above a wall which allows to
numerically compute MFPTs for different ligand and receptor
distributions and flow parameters both for two-dimensional (2D) and
three-dimensional (3D) movements \cite{korn:06,korn:07a}. Due to the
complex geometry arising from the receptor and ligand distributions
and the complexity of the position-dependent mobility functions
arising from the hydrodynamic equations, exact analytic results for
the MFPT cannot be obtained in this general case.

In this letter we present analytical advances for 2D motion with homogeneous
ligand coverage on the substrate.  In our model system, motion can occur only
in x- and z-directions and rotation is restricted about the y-axis. Due to the
assumption of homogeneous ligand coverage, motion in the x-direction is not
relevant. Thus we deal with two degrees of freedom, falling in z-direction and
rotation about the y-axis. This is the simplest model system which combines
rotational and translational degrees of freedom in a non-trivial
manner. Because cell movement usually occurs in the regime of small Reynolds
number, their coupling is determined by the Stokes equation for viscous
flow. In addition, we account for Brownian motion which is ubiquitous in
biological systems on the nanoscale and essential for receptor-ligand binding
to occur.  Here we show that this model system can be further reduced by
effectively integrating out the translational degree of freedom.  This results
in an ordinary differential equation for the MFPT of the rotational degree of
freedom which includes a non-trivial reaction term that represents the falling
motion of the cell. We show that this equation can be solved
analytically. From this solution we then derive various expressions that
describe asymptotic limits. By comparing with computer simulations, we finally
show for which parameter range our analytical results are valid. Moreover,
our analytical calculations still explain the numerical results outside this
parameter range when the effective reaction rate is used as a fit parameter.

\section{Model definition}

\begin{figure}
  \begin{center}
  \onefigure{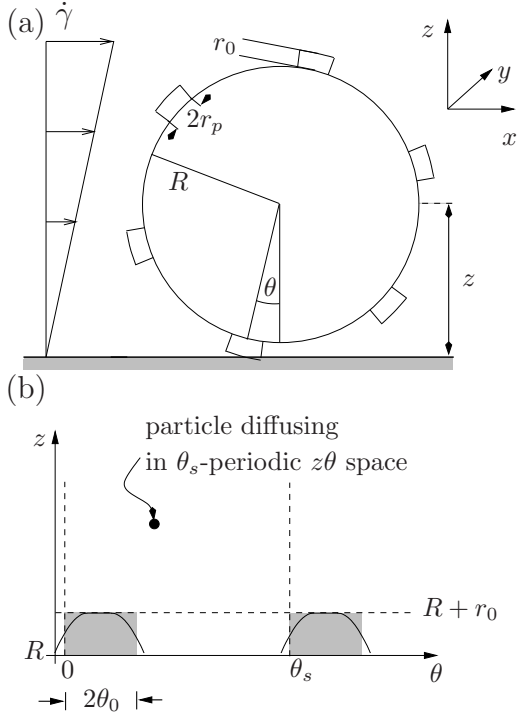}
  \caption{(a) A sphere (radius $R$) in linear shear flow (shear rate $\dot \gamma$)
    covered with $N_r = 6$ receptor patches with
   height $r_0$ and radius $r_p$ is shown. The sphere translates
   in the $xz$-plane. The orientation about the symmetry axis is given
   by the angle $\theta$.   
   (b) The encounter problem can be mapped to that of a particle moving in
   $z\theta$ space, where the motion in $\theta$-direction is $\theta_s :=
   2\pi/N_r$ periodic as the receptors are equidistantly distributed. The
   solid curves shown enclose the areas in which receptor-ligand encounter
   occurs.  For the calculation we approximate these areas by a rectangular
   area (grey shaded) of height $r_0$ and width $2\theta_0$.}
  \label{fig:setup}
  \end{center}
\end{figure}

We consider a sphere of radius $R$ moving in linear shear flow with
shear rate $\dot \gamma$ above a planar wall. As explained above, we
restrict its motion to two dimensions, that is the translational
motion of the sphere is restricted to a plane perpendicular to the
boundary wall, i.\,e. the $xz$-plane, and rotations are only allowed
about the $y$ axis (see \fig{fig:setup}a). As depicted in
\fig{fig:setup}a the circumference of the sphere lying in the
$xz$-plane is covered with $N_r$ equidistantly distributed receptor
patches of height $r_0 \ll R$ and radius $r_p \ll R$. The boundary
wall is homogeneously covered with ligands.  In order to drive it onto
the substrate, in vertical direction the sphere is subject to a
constant force $- F_z$. In experiments, this force arises from gravity
because cells or micro-beads are usually slightly denser than the
surrounding medium.  A receptor-ligand encounter occurs with certainty
whenever a receptor patch has some overlap with the boundary wall.
Because we also consider Brownian motion, the receptor-ligand
encounter is stochastic. Our goal is to calculate the corresponding
MFPT.  For the calculation of first passage times
the motion in $x$ direction can be neglected as the system is
translationally invariant in this direction due to homogeneous ligand
coverage.  Moreover, the regular distribution of receptor patches
generates a $\theta_s := 2\pi/N_r$ symmetry for the $\theta$
coordinate (i.\,e., the angle about the $y$-axis).  Therefore, we deal
with the situation illustrated in \fig{fig:setup}b of a diffusive
particle moving in the $\theta_s$-periodic $(z\theta)$ plane (i.e., a
cylindrical surface with $z > R$) with absorbing boundaries
$\theta(z)$ (solid lines in \fig{fig:setup}b) representing the
boundaries of the encounter areas.

If $T(z',\theta')$ is the MFPT to reach a point on the absorbing
boundary when started at some point $(z',\theta')$ then for practical
purposes only $\mean{T(z',\theta')}_{\theta'}$, i.\,e., the MFPT
averaged over all initial orientations, is a relevant quantity as it
is experimentally very difficulty to prepare a certain initial
orientation.  Concerning the initial height we previously showed
\cite{korn:07a} that for $z' > z_m > R + r_0$ the following relation
holds true
\begin{align}
  \label{eq:add}
  \mean{T(z',\theta')}_{\theta'} = T(z_m|z') + \mean{T(z_m,\theta_m)}_{\theta_m},
\end{align} 
where $T(z_m|z')$ is the mean time to fall from the initial height
$z'$ to the intermediate height $z_m$ and $\theta_m$ is the sphere's
orientation at height $z_m$. Eq.~(\ref{eq:add}) states that if the
angle averaged MFPT is known for some height $z_m$ the MFPT can be
calculated for any other height $z > z_m$ without further considering
the orientational degree of freedom. Moreover $T(z_m|z')$ can be
calculated exactly \cite{korn:07a}.  Thus for the following we choose
as the initial height of the disk always $z' = R + r_0$ and write
$T(\theta') := T(z'=R+r_0,\theta')$.

\section{Model reduction to one dimension}

In order to calculate $\mean{T(\theta')}_{\theta'}$ one must in
principle consider the motion in the full $(z\theta)$-plane. Here, we
show that under certain conditions a good approximation formula for
this MFPT can be derived by considering only the motion of a particle
in a periodic one-dimensional $\theta$-space. For that we first
approximate the area in the $(z\theta)$-plane in which encounter
occurs by a rectangle of width $2\theta_0$ (see \fig{fig:setup}b). As
an appropriate $\theta_0$ we choose the mean width of the encounter
interval the particle sees while being at height $z < R + r_0$, that
is
\begin{align}
  \theta_0 = \frac{1}{1 - e^{-F_z r_0/k_B T_a}}\int_R^{R + r_0} \de z \theta_0(z) \Psi_s(z), 
\end{align}
where
\begin{equation}
\label{eq:psisz}
\Psi_s(z) = \frac{F_z}{k_B T_a} e^{-F_z(z-R)/k_B T_a}
\end{equation}
is the stationary probability distribution for the height of the
sphere $z$ with Boltzmann's constant $k_B$ and ambient temperature
$T_a$. $\theta_0(z)$ is the boundary of the encounter area and depends
on receptor height and radius through $\theta_0(z) = \arccos(z/(R +
r_0))+ r_p$ \cite{korn:07a}. Thus $\theta_0$ increases with $r_p$.

Changes in $\theta$ are due to rotational diffusion with diffusion coefficient
$D_\theta$ and a drift $A_\theta \propto \dot\gamma$ arising from the
linear shear flow. This suggests to take the corresponding Fokker-Planck equation
as a starting point for a reduced model. In order to account for the effect
of the motion in z-direction, we argue that for $\theta \in [0, 2\theta_0]$ 
an encounter occurs only with a finite probability. 
Thus, for the probability $p(\theta,t)$ to have the orientation $\theta$ at
time $t$ we setup the following reaction-advection-diffusion equation for the interval
$[0,\theta_s[$
\begin{equation}
  \label{eq:probability}
  \partial_t p(\theta, t) = 
  D_\theta \partial_\theta^2 p(\theta,t) - 
  A_\theta \partial_\theta p(\theta, t)
  -\bar k(\theta)  p(\theta, t).
\end{equation} 
The reaction term is defined by $\bar k (\theta) = \bar k$ for $\theta
\in [0, 2\theta_0]$ and $\bar k (\theta) = 0$ otherwise. It accounts
for the finite probability that encounter occurs while the particle is
passing the interval $[0, 2\theta_0]$. This approximation makes sense
if $\theta_0$ is small and if in addition $\theta_0/\theta_s \ll 1$.
In our context, this assumption is justified because we have the
separation of length scales $r_p,r_0 \ll R$.  Then the time for
passing the interval $[0, 2\theta_0]$ is short compared to the time
between successive receptor-ligand encounter and the correlation
between the heights $z$ for successive encounters are small, thus
justifying our mean field treatment for the translational degree of
freedom. The assumption $\theta_0/\theta_s \ll 1$ also suggests to
take the limit $\theta_0 \rightarrow 0$ while keeping $2\theta_0 \bar
k =: k$ constant.  Then the term $-\bar k(\theta) p(\theta, t)$ in
\eq{eq:probability} becomes $-k\delta(\theta)p(\theta, t)$.
\eq{eq:probability} is a differential equation for the rotational
motion in which the effect of vertical motion has been absorbed into
the new model parameter $k$.

\section{Boundary conditions}

Because of the $\theta_s$-periodicity the probability $p(\theta, t)$
must fulfil periodic boundary conditions, i.\,e.  $p(\theta + \theta_s,t) =
p(\theta, t)$.  Furthermore, integrating \eq{eq:probability} over a
full period $\theta_s$ we obtain for the total loss of probability
\begin{align}
  \label{eq:bc1}
  \frac{d}{dt}\int_{\tilde \theta}^{\tilde \theta + \theta_s} d\theta p(\theta,t)
  = -k p(0,t),
\end{align}
i.\,e. $k$ denotes the rate of absorption at the boundary. In the limit
$k \rightarrow \infty$ we have purely absorbing boundaries with $p(0,t)
= p(\theta_s,t) = 0$. On the other hand integrating \eq{eq:probability}
over the open interval $]0,\theta_s[$ we obtain 
\begin{equation}
  \label{eq:bc2}
  \frac{d}{dt}\lim_{\epsilon \rightarrow 0}
  \int_{0+\epsilon}^{\theta_s - \epsilon} d\theta p(\theta,t)
  = J(0+,t) - J(\theta_s-,t) 
\end{equation}
with the probability current $J(\theta,t) := -(D_\theta \partial_\theta 
- A_\theta) p(\theta,t)$. 
Combining \eq{eq:bc1} and \eq{eq:bc2} we see that the reactive delta-function at 
$\theta = 0$ is equivalent to so-called \textit{radiation boundaries} \cite{szabo:80}
\begin{align}
  \label{eq:rad_bc}
  J(\theta_s,t) - J(0,t) = k p(0,t). 
\end{align}
In the limit of zero encounter probability, i.\,e. $k = 0$, the
probability flux leaving the interval on the right boundary is equal
to that entering the interval at the left boundary and the total
probability is conserved.

\section{Mean first passage time}

Let $G(\theta',t)$ denote the survival probability at time $t$,
i.e. the probability that no encounter has occurred until time $t$
under the condition that the initial orientation of the particle at
time $t' = 0$ was $\theta'$. With $p(\theta,t|\theta',0)$ being the
conditional probability for the particle of having the orientation $\theta$ at time
$t$ when the initial orientation of the particle at $t' = 0$ was
$\theta'$, the survival probability can be written as $G(\theta',t) =
\int_0^{\theta_s}\de \theta p(\theta,t|\theta',0)$.
$G(\theta',t)$ obeys the adjoint equation \cite{honerkamp:94}
\begin{align}
  \label{adjoint}
  \partial_t G(\theta',t) = 
  (D_\theta \partial_{\theta'}^2 + A_\theta \partial_{\theta'} - k\delta(\theta'))
  G(\theta',t).
\end{align}
The mean first passage time $T(\theta')$ follows from the survival probability as
$T(\theta') = \int_0^\infty\de t G(\theta',t)$. 
Thus, an equation for the MFPT is obtained by integrating \eq{adjoint} over all times.
With $\lim_{t\rightarrow\infty}G(\theta',t) = 0$ and $G(\theta',0) = 1$
the MFPT is the solution of \cite{honerkamp:94} 
\begin{align}
  \label{eq:mfpt}
   (D_\theta \partial_{\theta'}^2 + A_\theta \partial_{\theta'} - k\delta(\theta'))
  T(\theta') = -1.
\end{align}
The general solution of \eq{eq:mfpt} is given by
\begin{align}
  \label{eq:gensol}
  T(\theta') = -\frac{\theta'}{A_\theta} + \frac{D_\theta}{A_\theta^2}
  \left(1 - a_1 e^{-\frac{A_\theta}{D_\theta}\theta'}\right) + a_2,
\end{align}
with two integration constants $a_1,a_2$ which have to be determined
in order to match the boundary conditions, i.e. periodicity $T(0) =
T(\theta_s)$. The condition corresponding to \eq{eq:rad_bc} is
$\partial_{\theta'}T|_{\theta' = 0} - \partial_{\theta'}T|_{\theta'
=\theta_s} = (k/D_\theta) T(0)$, which follows from integrating
\eq{eq:mfpt} from $-\epsilon$ to $\epsilon$ with some $\epsilon > 0$
and then taking the limit $\epsilon \rightarrow 0$. This gives
\begin{align}
  \label{eq:thetadep}
  T(\theta') = \frac{\theta_s}{A_\theta}
  \frac{1 - e^{-\frac{A_\theta}{D_\theta}\theta'}}
       {1 - e^{-\frac{A_\theta}{D_\theta}\theta_s}} 
- \frac{\theta'}{A_\theta} + \frac{\theta_s}{k}.
\end{align}
This equation has to be corrected by the physical expectation
that $T(\theta') = 0$ for
$\theta' \in [0,2\theta_0]$ as we used for the initial height always
$z' = R + r_0$. Therefore, for these orientations encounter occurs
instantaneously.  This problem is fixed by replacing $\theta_s$ by
$\Delta \theta := \theta_s - 2\theta_0$ in \eq{eq:thetadep}. Then
averaging over all initial orientations including some for which $T$
is zero we arrive at our central result
\begin{align}
  \label{eq:fullres}
  \mean{T(\theta')}_{\theta'} 
  &= \frac{1}{\theta_s}\int_{0}^{\Delta\theta}\de \theta' T(\theta')\nonumber\\
  &= \frac{A_\theta\Delta\theta^2\coth\left(\frac{A_\theta\Delta\theta}{2D_\theta}\right)
    - 2D_\theta\Delta_\theta}
       {2A_\theta^2\theta_s} + \frac{\Delta \theta^2}{k\theta_s}.
\end{align}
As one would expect $\mean{T(\theta')}_{\theta'}$ becomes the smaller
the larger the encounter probability, i.e. the larger the
reaction rate $k$.

\section{Parameter estimates}

In order to apply \eq{eq:fullres} to the situation of a particle moving in the
$(z\theta)$-plane we still have to provide expressions for the reaction rate $k$
as well as for the diffusion constant $D_\theta$ and the drift $A_\theta$.
Regarding the reaction rate, we have to consider the full system
again. We first note that in
the stationary state the probability for the sphere to be at a height between
$z$ and $z+\de z$ is given by $\Psi_s(z)\de z$ following from \eq{eq:psisz}. 
Thus, we get for the
probability for encounter while the sphere is oriented such that $\theta \in
[0,2\theta_0]$ (valid in the limit $\theta_0 \ll 1$) $p_z := 1 - \exp(-F_z
r_0/k_B T_a)$.  On the other hand using the originally introduced rate $\bar
k$ the mean probability for encounter is $p_{\bar k} := 1 - \exp(-\bar k
\tau)$ where $\tau$ is the mean time it takes the particle to pass the
interval of length $2\theta_0$. For purely diffusive motion $A_\theta = 0$ we
estimate $\tau = \theta_0^2/2D_\theta$ which is the mean first passage time to
reach the boundary of $[0,2\theta_0]$ when initially started at $\theta(t = 0)
= \theta_0$. In the limit of very large drift motion becomes purely
deterministic and we expect $\tau = 2\theta_0/A_\theta$. These two limiting
cases may be combined to provide $1/\tau = 2 D_\theta/\theta_0^2 +
A_\theta/2\theta_0$.  Under the assumption that the position between two
successive approaches of the interval $[0,2\theta_0]$ is independent one can
get an estimate for $k$ from the condition $p_{\bar k} = p_z$
\begin{align}
  \label{eq:rate}
  k = \frac{F_z r_0}{k_B T_a}\frac{2\theta_0}{\tau}
  =\frac{F_z r_0}{k_B T_a} \left(\frac{4D_\theta}{\theta_0} + A_\theta\right),
\end{align}
where we used $k = 2\theta_0 \bar k$. From \eq{eq:rate} one sees that
encounter becomes the more probable the larger the vertical drift
force $F_z$ and the faster rotation given by $D_\theta$ and
$A_\theta$.

For a sphere in linear shear flow above a wall the
coefficients $D_\theta$ and $A_\theta$ depend on the height $z$ as
\begin{align}
  \label{eq:zdep}
  D_\theta(z) = \frac{k_B T_a}{8 \pi \eta R^3}\tilde\beta^{rr}(R/z),\;\;
  A_\theta(z) = \frac{\dot\gamma}{2}(1-\tilde\beta^{dr}(R/z)),
\end{align}
with $\eta$ denoting the viscosity of the fluid.
$\tilde\beta^{rr}(1/z)$ and $\tilde\beta^{dr}(R/z)$ are dimensionless
functions including the $z$-dependence. A numerical scheme
that allows to accurately calculate these functions at arbitrary
heights can be found in \cite{jones:92,jones:98}. In order to
compare the results of the model presented here with the numerical
solutions of the MFPT problem we use mean-field values in the
following way
\begin{align}
  \label{eq:constants}
  D_\theta := \int\limits_R^\infty\de z \Psi_s(z) D_\theta(z),\;\; 
  A_\theta := \int\limits_R^\infty\de z \Psi_s(z) A_\theta(z).
\end{align}
That is we take for $D_\theta$ and $A_\theta$ averages with respect to
the stationary probability function $\Psi_s(z)$ from \eq{eq:psisz} of
the $z$-dependent quantities $D_\theta(z),A_\theta(z)$ defined in
\eq{eq:zdep}.

\section{Comparison to simulation results}

\begin{figure}
  \begin{center}
  \resizebox{.9\linewidth}{!}{\includegraphics{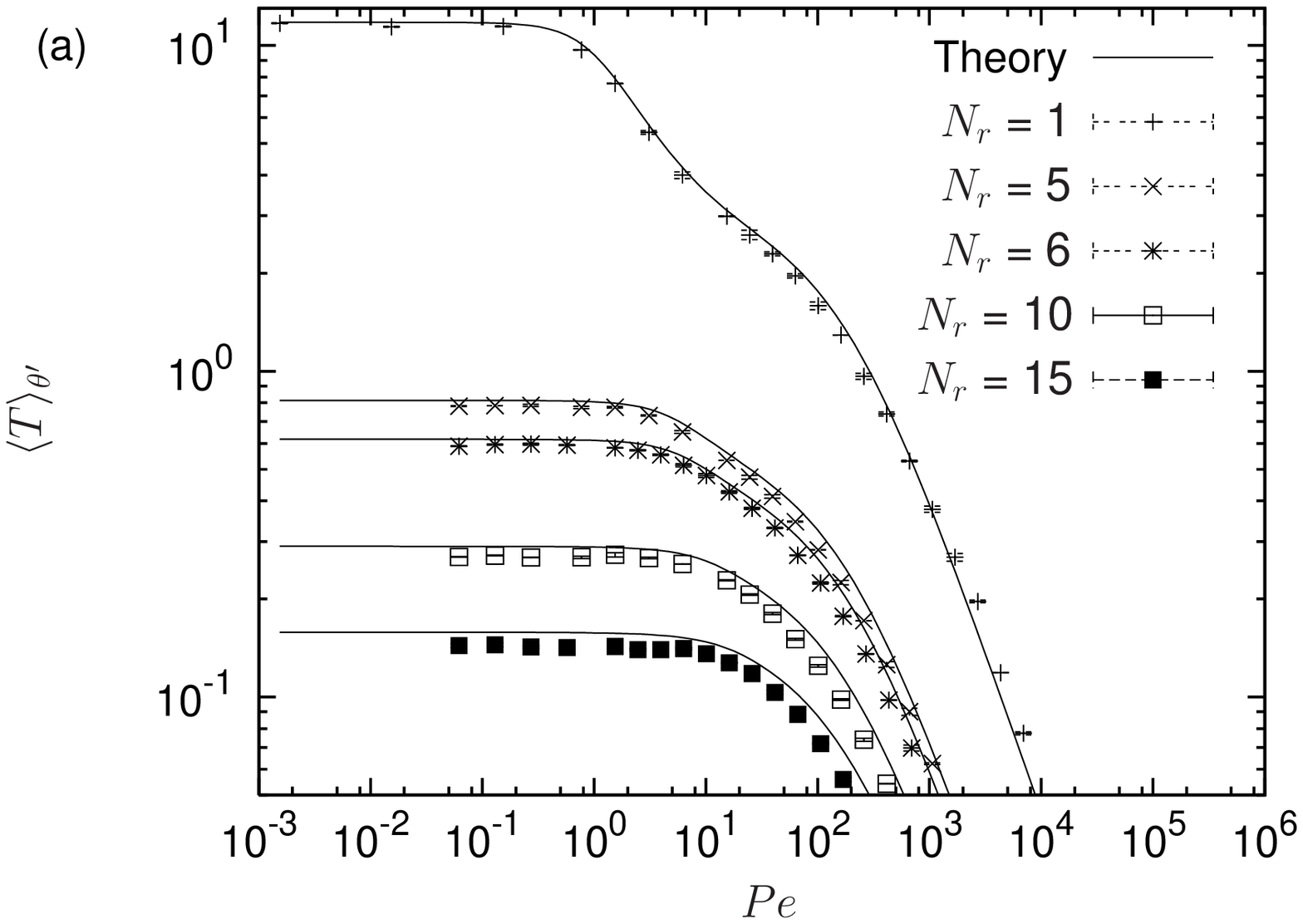}}
  \resizebox{.9\linewidth}{!}{\includegraphics{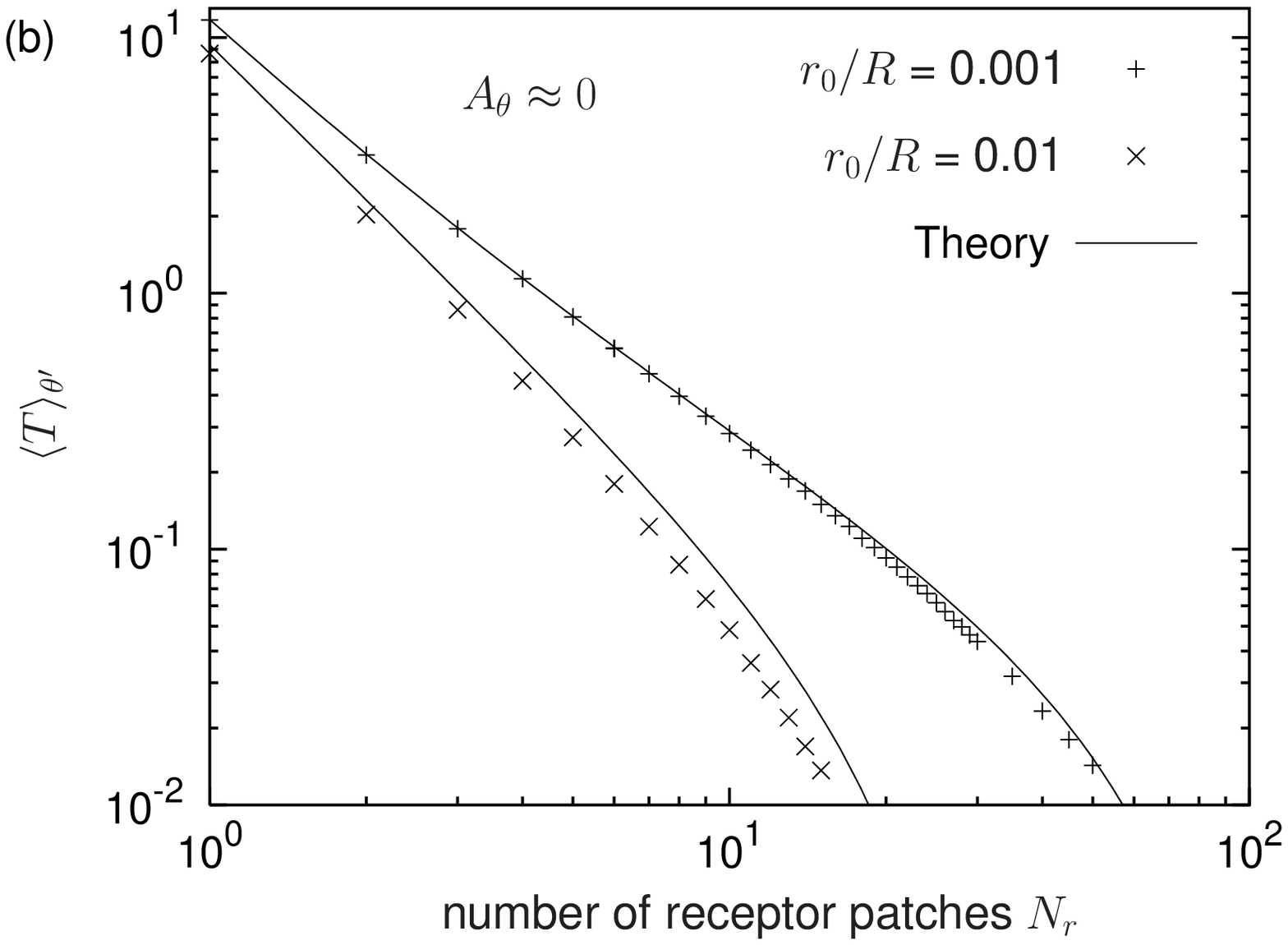}}
  \caption{Comparison between numerical results for the MFPT
    $\mean{T}_{\theta'}$ averaged over all initial orientations and the
    mean-field approximation according to
    eqs.~(\ref{eq:fullres},\ref{eq:rate},\ref{eq:constants}).  The
    numerical results where obtained as described in
    ref.~\cite{korn:07a} for an initial height $z = 2R$. In order to
    display the MFPT with initial height $z = R+r_0$ the falling time
    $T(R+r_0|2R)$ was subtracted according to \eq{eq:add}.  (a)
    $\mean{T}_{\theta'}$ as a function of the dimensionless P{\'e}clet
    number $Pe$ for different numbers of receptor patches $N_r$. 
    (b) $\mean{T}_{\theta'}$ as a function of $N_r$ for two different 
    values of the patch height $r_0$.
    (Other parameters: patch radius $r_p = 0.001 R$, vertical drift $F_z = Pe_z R/k_B T_a$ with 
    $Pe_z = 50$.)}
  \label{fig:pedep}
  \end{center}
\end{figure}
In \fig{fig:pedep} we compare our results for the angle averaged MFPT
$\mean{T}_{\theta'}$ as obtained from the analytical calculation
(solid lines) and from computer simulations (symbols). For this
purpose we rescale time in units of the diffusive time scale $6\pi\eta
R^3/k_B T_a$. In \fig{fig:pedep}a $\mean{T}_{\theta'}$ is shown as a
function of the P{\'e}clet number $Pe := 6\pi\eta R^3\dot\gamma/k_B
T_a$ (other parameters are defined in the figure caption). The
P{\'e}clet number is a dimensionless measure for the relative
importance of deterministic and diffusive motion and $A_\theta \propto
Pe$.  In the case of zero rotational drift, $A_\theta = 0$, $Pe = 0$
and motion is purely diffusive.  In addition to the simulation results
(symbols) \fig{fig:pedep}a also shows our main result (lines)
\eq{eq:fullres} where we used eqs.~(\ref{eq:rate}) and
(\ref{eq:constants}) for the rate $k$ and the constants $A_\theta,
D_\theta$, respectively.  The agreement between the mean-field and
simulation results is surprisingly good for the parameter values
chosen. In particular, even the small shoulder for $N_r = 1$ seems to
be reproduced by the analytical result.  In \fig{fig:pedep}b
$\mean{T}_{\theta'}$ is shown as a function of the number of receptor
patches $N_r$ in the diffusive limit $A_\theta = 0$ and for two
different values of the patch height $r_0$.  Whereas the theoretical
approximation with the rate given by \eq{eq:rate} works quite well for
$r_0 = 10^{-3}R$, for a tenfold larger $r_0$ we notice clear
deviations. The larger $r_0$ the larger $\theta_0$ and the assumption
of the dwelling time within $[0,2\theta_0]$ being small is less valid.
Nevertheless, \eq{eq:fullres} describes the functional dependence of
$\mean{T(N_r)}_{\theta'}$ in a qualitative way. In fact quantitative
agreement can be achieved by using the reaction rate $k$ as a fit
parameter (not shown).

\section{Asymptotic limits}

\begin{figure}
  \resizebox{.9\linewidth}{!}{\includegraphics{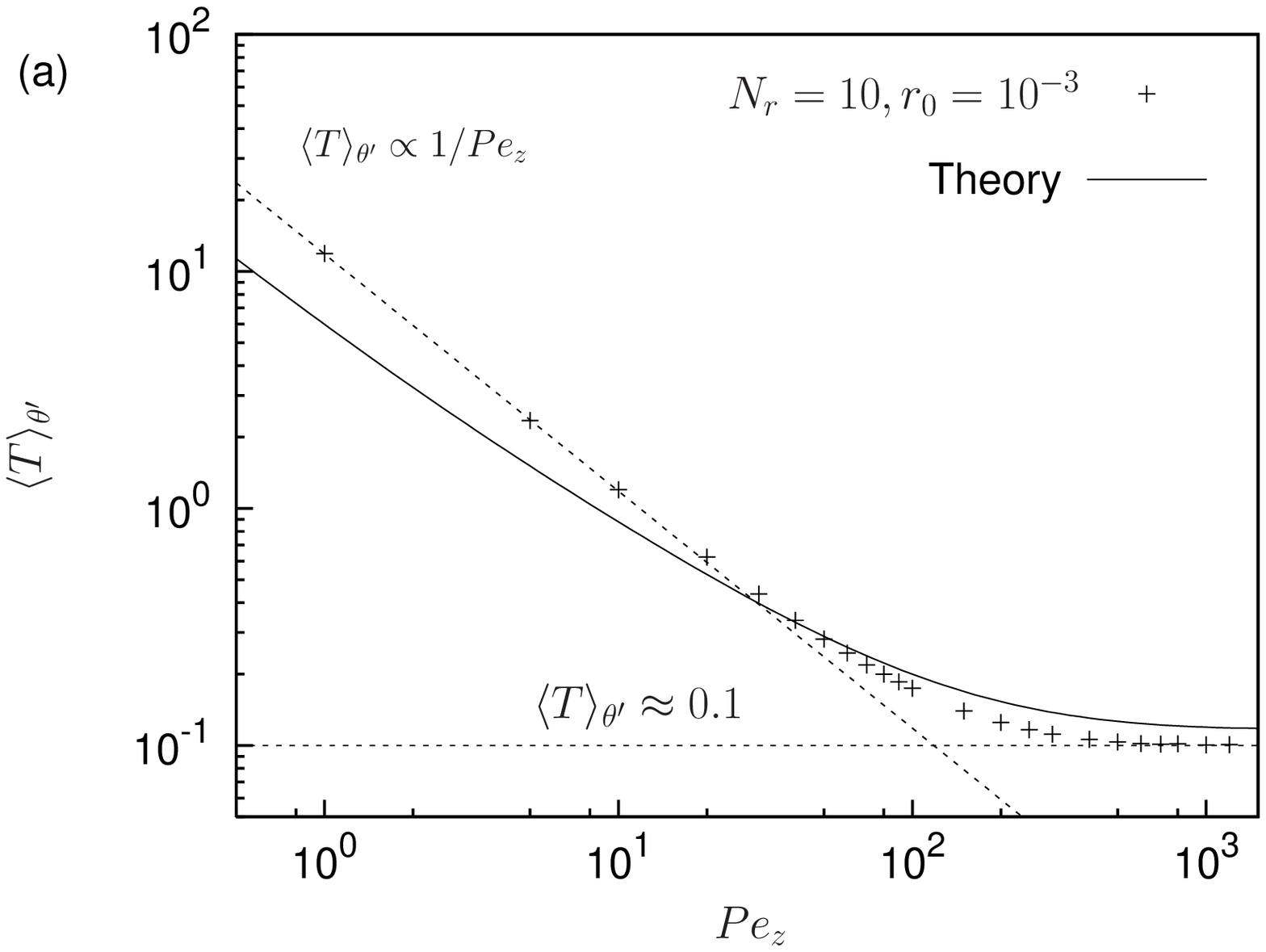}}\\
  \resizebox{.9\linewidth}{!}{\includegraphics{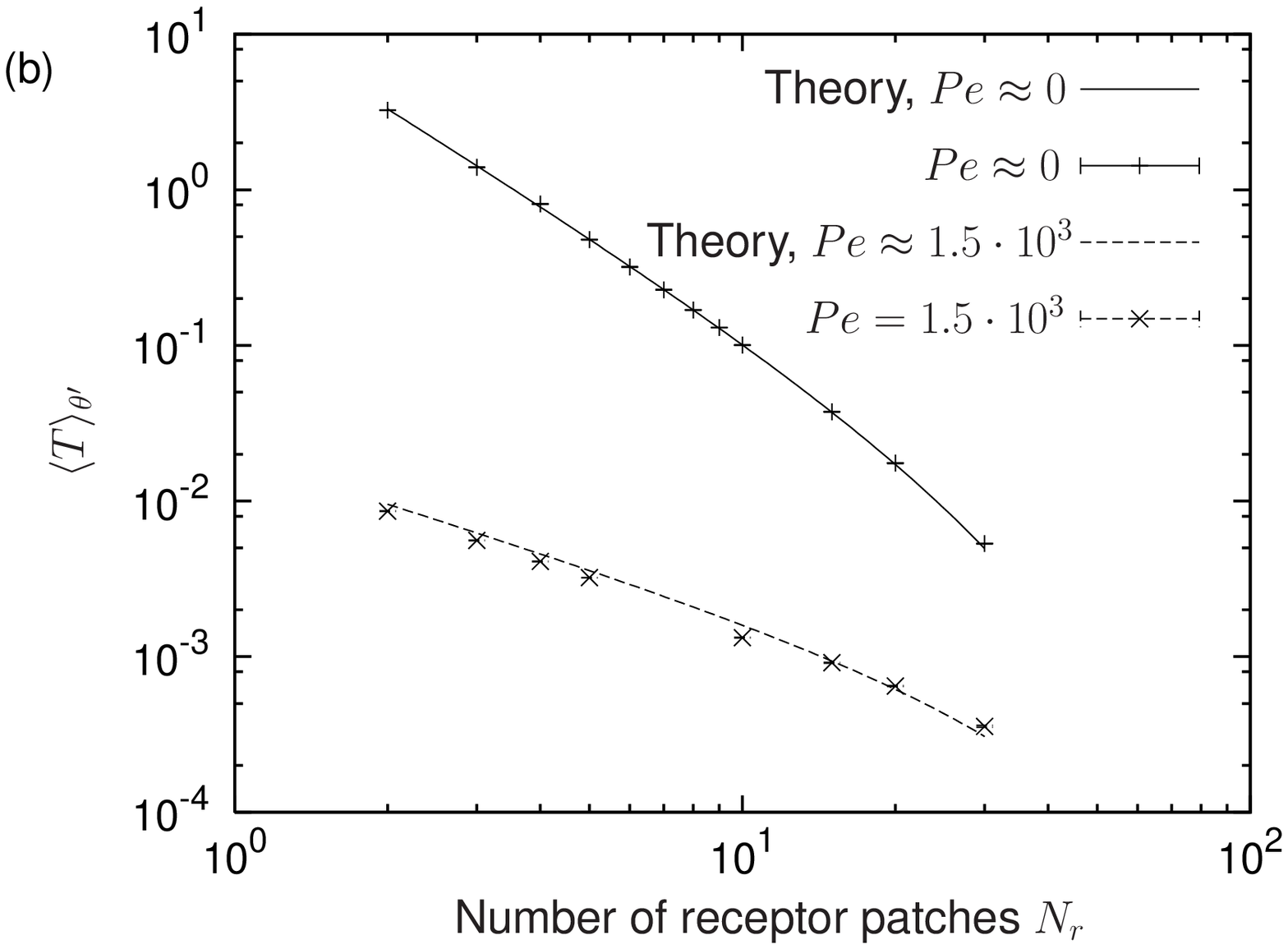}}
  \caption{(a) The numerically obtained MFPT $\mean{T}_{\theta'}$ as a function 
    of the strength of the vertical drift $Pe_z$ (+). The dashed lines show the
    asymptotic behaviour $\mean{T}_{\theta'} \propto 1/Pe_z$ for small $Pe_z$
    and the value of $\mean{T}_{\theta'}$ around $Pe_z = 10^3$.
    The solid line shows the mean-field approximation according to
    eqs.~(\ref{eq:fullres},\ref{eq:rate},\ref{eq:constants}).
    ($N_r = 10$, $r_0=r_p=10^{-3}R$). (b) The different scaling behaviour
    of $\mean{T}_{\theta'}$ in regard to the number of receptor patches $N_r$
    for zero and large rotational drift $A_\theta$ at $Pe_z = 10^3$ is shown.
    The theoretical result from \eq{eq:fullres} matches well
    the numerical results (symbols) when for $k$ infinity and $k = 2A_\theta$ is chosen
    for $Pe = 0$ and $Pe \approx 1500$, respectively ($r_0 = r_p = 10^{-3}R$).}
  \label{fig:asym}
\end{figure}

The results shown in \fig{fig:pedep} were obtained for a fixed
vertical drift with strength given by $Pe_z = 50$. Here, $Pe_z := F_z
R/k_B T_a$ is the P{\'e}clet number for the $z$-direction.
Fig.~\ref{fig:asym}a displays the comparison between the numerical
result of $\mean{T}_{\theta'}$ and the approximation \eq{eq:fullres}
as a function of $Pe_z$ in the diffusive limit $A_\theta \approx 0$.
For small $Pe_z$ the numerical results for the MFPT shows a $1/Pe_z$
scaling behaviour as indicated by the dashed line in \fig{fig:asym}a.
For larger $Pe_z$, $\mean{T}_{\theta'}$ plateaus.  At even larger
$Pe_z$ the MFPT increases with increasing $Pe_z$ (not shown) as
$D_\theta(z) \rightarrow 0$ for $z\rightarrow R$ \cite{jones:98}.  One
sees that the approximation \eq{eq:fullres} underestimates the
numerical result for small values of $Pe_z$ and overestimates it for
large $Pe_z$.  For intermediate values of $Pe_z$ of the order of
10-100 we find good agreement between the two results as demonstrated
before in \fig{fig:pedep}. For large $Pe_z \approx 10^3$ the numerical
result provides $\mean{T}_{\theta'}\approx 0.1$. About the same value
is given by the first term in \eq{eq:fullres}. This means that in the
diffusive limit and for large $Pe_z$ an encounter occurs almost with
probability one for $\theta \in [0,2\theta_0]$. This is plausible as
then the duration time for $\theta \in [0,2\theta_0]$ is long enough
such that the particle will most probably encounter a height $z < R +
r_0$ while the receptor patch points downwards. Therefore, the rate
$k$ for $Pe_z \approx 10^3$ is rather infinity than the value given by
the estimate \eq{eq:rate}.  At small values of $Pe_z$ we expect
combining \eq{eq:fullres} and \eq{eq:rate} the second term in
\eq{eq:fullres} to be dominant and $\mean{T}_{\theta'} \propto
1/(D_\theta Pe_z)$.  As the diffusion coefficient $D_\theta(z)$ is a
monotonically increasing with increasing distance $z$ from the wall
also the term $D_\theta$ defined in \eq{eq:constants} becomes larger
with decreasing $Pe_z$.  This is the reason why the estimate
\eq{eq:fullres} for $\mean{T}_{\theta'}$ does not provide the right
scaling behaviour. On the other hand using $D_\theta|_{Pe_z = 50}$
also at larger $Pe_z$ values a much better agreement (not shown)
between the theoretical estimate and the numerical result of the MFPT
is obtained. That implies that the faster rotations far away from the
wall, which are included in the definition \eq{eq:constants}, are not
relevant for the encounter process which happens only very close to
the wall.
 
In \fig{fig:asym}b we demonstrate that \eq{eq:fullres} matches well to
the numerically obtained MFPT for $A_\theta = 0$ and large $Pe_z$ when
$k$ is set to infinity as discussed above. There, for $Pe_z = 10^3$,
the numerical result for $\mean{T}_{\theta'}$ (+) and the theoretical
result \eq{eq:fullres} (full line) are shown.  For the dependence on
the number of receptor patches $N_r$ we find in this limit (and for
$N_r$ small)
\begin{align}
  \label{eq:approx1}
  \mean{T}_{\theta'} \approx \frac{\Delta \theta^3}{12\theta_s D_\theta} \approx
  \frac{4\pi^2}{12 D_\theta N_r^2},
\end{align}
i.e. the MFPT approximately scales as $1/N_r^2$. 
The situation is different in the deterministic limit, i.e. for large $A_\theta$.
Then, \eq{eq:fullres} is approximately 
\begin{align} 
  \label{eq:approx2}
  \mean{T}_{\theta'} \approx \frac{\Delta \theta^2}{\theta_s}
  \left(\frac{1}{2A_\theta} + \frac{1}{k}\right) \approx
  \frac{\pi}{N_r} \left(\frac{1}{2A_\theta} + \frac{1}{k}\right),
\end{align}
i.e. the MFPT scales as $1/N_r$ at small numbers of receptor patches $N_r$.
In \fig{fig:asym}b we also show a comparison between \eq{eq:approx2} (dashed
line) and the numerically obtained MFPT ($\times$).  Here we find that good agreement
is obtained when we choose $k\approx 2A_\theta$, which is of the same order of
magnitude as the estimate given by \eq{eq:rate}. Thus, in contrast to the
diffusive limit, the second and the first term in \eq{eq:fullres} are of the
same order.

\section{Summary and outlook}

In this letter we have derived an approximate expression
\eq{eq:fullres} for the angle averaged MFPT for receptor-ligand
encounter between a sphere equidistantly covered with receptor patches
and a wall homogeneously covered with ligands in a 2D geometry.  The
main idea of our analysis was to integrate out the motion in
$z$-direction by absorbing falling in a reaction term for the
rotational degree of freedom.  The coefficients for the diffusion and
drift terms of this equation were estimated from mean field arguments
in \eq{eq:constants}.  Our derivation was based on the central
assumption that the heights of the sphere at two successive times of a
receptor patch pointing downwards is uncorrelated. This is not true in
general, and accordingly the result derived for the reaction rate $k$
in \eq{eq:rate} is valid only for a small range of parameter values.
However, the rate $k$ can also be viewed as a fit parameter. In that
case \eq{eq:fullres} matches the results obtained in computer
simulations of the full problem over a large range of parameters.

In the future, our analysis could be extended in different ways.  Here
we have only considered a homogeneous ligand density.  For
non-homogeneous ligand coverage $\rho_l < 1$ not every receptor-wall
encounter is a productive receptor-ligand encounter. This might again
be expressed by an appropriate choice of the rate $k$ with $k\propto
\rho_l$. Even for homogeneous ligand density, not every
receptor-ligand encounter has to lead to functional adhesion under
flow. Depending on the receptor-ligand system under consideration,
additional steps might be required to achieve a stable bond.
Conceptually, one could regard the first encounter as formation of an
encounter complex \cite{berg:77}. The next step would then be the
transition into a final complex. In a such a two-state system, also
unbinding becomes important, as it characterizes the stability of the
final complex.  Again this process might be expressed by an
appropriate choice of the rate $k$ as used here. If complete unbinding
occurs, there is also the possibility that the cell forms a new bond
downstream of the old one.  At sufficient densities of receptors and
ligands, this mechanism eventually leads to the physiologically very
relevant process of rolling adhesion. Recently we have extended the
computer simulations used above to measure MFPTs to also simulate the
process of rolling adhesion \cite{korn:08}.  Future work has to show
how these simulations can now be made more efficient using the
dimensional reduction introduced here.  In general, analytical
progress by dimensional reduction and introduction of appropriate
reaction terms might be a very promising strategy also for other
biological systems which involve a complex interplay between different
transport modes.

\begin{acknowledgments}
  This work was supported by the \textit{Center for Modelling and Simulation
  in the Biosciences} (BIOMS) and the Cluster of Excellence
  \textit{CellNetworks} at Heidelberg.
\end{acknowledgments}


\begin{thebibliography}{10}
\expandafter\ifx\csname url\endcsname\relax\def\url#1{\texttt{#1}}\fi

\bibitem{alberts}
\Name{Alberts B., Johnson A., Lewis J., Raff M., Roberts K. \and Walter P.}
  \Book{Molecular Biology of the Cell} 5th Edition (Garland Science) 2008.

\bibitem{berg:77}
\Name{Berg H.~C. \and Purcell E.~M.} \REVIEW{Biophys. J.}{20}{1977}{193}.

\bibitem{seifert:05c}
\Name{Reister E. \and Seifert U.} \REVIEW{Europhys. Lett.}{71}{2005}{859}.

\bibitem{haenggi:02}
\Name{Goychuk I. \and Hanggi P.} \REVIEW{Proc. Natl. Acad. Sci. USA}{99}{2002}{3552}.

\bibitem{holcman:07}
\Name{Holcman D.} \REVIEW{J. Stat. Phys.}{127}{2007}{471}.

\bibitem{lawrence:91}
\Name{Lawrence M.~B. \and Springer T.~A.} \REVIEW{Cell}{65}{1991}{859}.

\bibitem{korn:06}
\Name{Korn C. \and Schwarz U.~S.} \REVIEW{Phys. Rev. Lett.}{97}{2006}{138103}.

\bibitem{alon:97}
\Name{Alon R., Chen S., Puri K.~D., Finger E.~B. \and Springer T.~A.}
  \REVIEW{J. Cell Biol.}{138}{1997}{1169}.

\bibitem{springer:01}
\Name{Chen S. \and Springer T.~A.} \REVIEW{Proc. Natl. Acad. Sci. USA}{98}{2001}{950}.

\bibitem{pierres:98}
\Name{Pierres A., Feracci H., Delmas V., Benoliel A.-M., Thiery J.-P. \and
  Bongrand P.} \REVIEW{Proc. Natl. Acad. Sci. USA}{95}{1998}{9256}.

\bibitem{hammer:00a}
\Name{Greenberg A.~W., Brunk D.~K. \and Hammer D.~A.} \REVIEW{Biophy.
  J.}{79}{2000}{2391}.

\bibitem{yago:07}
\Name{Yago T., Zarnitsyna V.~I., Klopocki A.~G., McEver R.~P. \and Zhu C.}
  \REVIEW{Biophys. J.}{92}{2007}{330}.

\bibitem{korn:07a}
\Name{Korn C.~B. \and Schwarz U.~S.} \REVIEW{J. Chem. Phys.}{126}{2007}{095103}.

\bibitem{szabo:80}
\Name{Szabo A., Schulten K. \and Schulten Z.} \REVIEW{J. Chem. Phys.}{72}{1980}{4350}.

\bibitem{honerkamp:94}
\Name{Honerkamp J.} \Book{Stochastic Dynamical Systems} (VCH Publishers, Inc.)
  1994.

\bibitem{jones:92}
\Name{Perkins G.~S. \and Jones R.~B.} \REVIEW{Physica A}{189}{1992}{447}.

\bibitem{jones:98}
\Name{Cichocki B. \and Jones R.~B.} \REVIEW{Physica A}{258}{1998}{273}.

\bibitem{korn:08}
\Name{Korn C. B. \and Schwarz U.~S.} \REVIEW{Phys. Rev. E}{77}{2008}{041904}.

\end{thebibliography}
\end{document}